\documentclass[12pt,a4paper]{article}
\usepackage{latexsym}
\usepackage[latin1]{inputenc}
\usepackage[dvips]{graphics}
\usepackage{bbm}
\usepackage{amssymb,amstext,amsfonts,amsmath}

\begin{document}




\begin{center}
{\Large  On some mathematical identities resulting from evaluation of the partition function for an electron moving in a periodic lattice. \\  }
\vspace{0.5cm}
Jakub J\c{e}drak \\ 
Marian Smoluchowski Institute of Physics, Jagiellonian University, \\
Reymonta 4, 30-059 Krak\'{o}w, Poland \\
e-mail: jedrak@th.if.uj.edu.pl
\end{center}

\begin{abstract}
We consider a simple model of the dynamics of a single electron in a crystal lattice. Although this is a  standard problem in condensed matter physics, alternative ways of evaluating a partition function for such a system lead to equalities, that may be interesting from the point of view of mathematical analysis, combinatorics and graph theory.  
\end{abstract}

\section*{Introduction}

In this paper we present results which may be interesting from the point of view of pure mathematics, but originate from some properties of mathematical formalism used to describe certain physical systems. Namely, we derive some generalizations of the integral representation of the modified Bessel function $I_0\left(2 \xi \right)$,
\begin{equation}
 \frac{1}{2\pi}\int_{-\pi}^{\pi}e^{2\xi \cos (k)}dk = \sum_{\nu = 0}^{\infty}   \frac{\xi^{2 \nu}}{\nu!\nu!}, 
\label{once less bessel}
\end{equation}
here given in the form of the Taylor expansion around $\xi = 0$ (see \cite{AS}, 9.6.16, p. 376). 
The important observation is that formula (\ref{once less bessel}) describes a \textit{partition function} of the \textit{tight-binding model} of a single electron moving in an infinite, one-dimensional chain of atoms. This model and its generalizations are  widely used in condensed matter physics and may be found in any standard textbook on that field, (see e.g. \cite{Fazekas}, p. 11).

If we consider, in the framework of the same model, some $D$-dimensional crystal lattice, and if we write down the corresponding partition function, we obtain equality analogous to (\ref{once less bessel}). Namely, its l.h.s is then a sum or integral over some subset of $ \mathbb{R}^D$, having nontrivial symmetry properties. 
On the other hand, the r.h.s. of both formula (\ref{once less bessel}) and any of its generalizations is a series in one or several complex variables, and can be found with the help of the method described in detail in the present paper. In that way, we are able to establish several identities, which may be interesting from the point of view of mathematical analysis. What is more important, our method involves  some basic techniques and concepts of combinatorics and graph theory, therefore it may be interesting from the point of view those disciplines.

The results presented here  have been obtained with the help of a standard mathematical  formalism and techniques used in theoretical condensed-matter physics. This fact has two important consequences. First, our notation is generally an usual one for that field. 
This results in occurrence of additional numerical factors and indices, which may  be found redundant, in the  formulas presented. Hence, throughout the paper we use units in which Planck constant $\hbar$ and Boltzmann constant $k_B$ are equal unity, $\hbar = k_B = 1$. 

Second, our results may be found  not rigorous enough from the point of view of pure mathematics.  Nonetheless, we believe that they are correct, and they can be given a more formal shape if necessary. This should be one of the subjects of our future investigation.
 
This paper is organized as follows: in section 1. we introduce some concepts necessary to understand our method and we describe the method itself. In section 2. we  analyze various lattice geometries and present the results obtained. Section 3. consists of appendices, where some details, omitted in the main text, are presented. 
\section{The method}


\subsection{Preliminaries: physical context}


Let us consider a  motion of a single electron in a periodic crystal lattice. 
To construct mathematical model of such a physical system, we have to provide a description of a crystal lattice, as well as to express the  quantum dynamics of the  electron on that lattice.

To model a crystal lattice, we define first the  $D$-dimensional \textit{abstract lattice}, $\mathbf{\Lambda}$, in a standard manner, as a set of \textit{lattice vectors} $\mathbf{R}$,
$\mathbf{r} \in\mathbb{R}^D$, being the  linear combinations with integral coefficients of  $D$ fundamental translation vectors $ \mathbf{a}_{1},  \mathbf{a}_{2},\ldots, \mathbf{a}_{D}$ 

\begin{equation}
\mathbf{R}  = \sum_{p=1}^{D} n_{p} \mathbf{a}_{p}.
\label{lattice}
\end{equation}
However, appropriate model of the real crystal structure is constructed by attaching basis of $M \geq 1$ atoms to each  point of the abstract lattice (see e.g.  \cite{Kittel}, p. 5). We denote positions of atoms by vectors $\mathbf{i}$, called
\textit{lattice sites}, each lattice site has its  corresponding lattice vector, $\mathbf{R(\mathbf{i})}$. For lattices  with $M=1$,  lattice vectors  can be chosen in such a way that they coincide with lattice sites. This is the case of \textit{Bravais  lattice},  and most, although not all, cases analyzed in this paper belong to this class. On the other hand non-Bravais  lattices consist of $M>1$ pairwise disjoint \textit{sublattices}, with any two points of a given sublattice being equivalent. 

It is of crucial importance not to confuse  $\mathbf{\Lambda}$ with the set of lattice sites  $\mathbf{i}$, denoted $\mathbf{\Xi}$. From now on we use the term  \textit{ lattice} only for
$\mathbf{\Xi}$, whereas $\mathbf{\Lambda}$ is always termed \textit{abstract lattice}, as 
before.
$\Lambda = |\mathbf{\Lambda}|$  ($\Xi = |\mathbf{\Xi}|$) is the number of the lattice vectors (sites), respectively. Obviously,  $\Xi = M \Lambda$, and this number may be finite or not. I the former case the periodic boundary conditions (PBC) have to be imposed to ensure periodicity. 

Apart from abstract lattice $\mathbf{\Lambda}$ we have to remind the Reader also  the notion of the  \textit{reciprocal lattice}, defined as the set of  vectors $\mathbf{K}$ having the following property (see e.g. \cite{Kittel}, p. 61),
\begin{equation}
 e^{i \mathbf{K}\cdot \mathbf{R}} = 1,   ~~~~~\text{for any}  ~~~ \mathbf{R}\in \mathbf{\Lambda} .
\end{equation}

Now we return to the question of a quantum-mechanical description of the dynamics of the electron.
To be able to give such a  description, first we have to specify the Hilbert  space $\mathcal{H}$ of the state vectors, and then choose some basis vector set in this space.  It turns out that
set $\mathbf{\Xi}$ in a real space provides the convenient basis in $\mathcal{H}$.
Namely, we define state   $ |\mathbf{i} \rangle \ \in \mathcal{H}$ 
as describing the electron cloud centered on the lattice site $\mathbf{i}$. This is so called \textit{Wannier basis} or \textit{Wannier representation} 
\footnote{It may be also regarded as some kind of \textit{coordinate representation}
 of quantum mechanics, but in the discretized, rather then continuous, real space. We use those three terms interchangeably in the following sections.}.
Such defined states form a complete and orthonormal basis set,
\begin{equation}
\sum_{\mathbf{i} \in \mathbf{\Xi}} |\mathbf{i} \rangle \langle \mathbf{i}| = \hat{\mathbf{1}}, ~~~~~~
 \langle \mathbf{i}|\mathbf{j} \rangle \ = \delta_{\mathbf{i}\mathbf{j}}.
\label{Kronecker delta for Wannier}
\end{equation}
It is convenient to treat our Hilbert space as a subspace of the larger one, called the \textit{Fock space}, $\mathcal{F}$. This new space, apart from the just defined one-particle states $ |\mathbf{i} \rangle \ $, consist of other $N$-particle states, with $N = 0, 2, \ldots, N_{max}$, $N_{max} = \Xi$, and may be written as a direct sum of its $N $-particle sectors, $\mathcal{F} = \mathcal{F}_0 \oplus \mathcal{F}_1 \oplus \cdots \oplus \mathcal{F}_{N_{max}}$. 
Although we are interested in one-particle sector only, $\mathcal{F}_1 \equiv \mathcal{H}$, we use the general formalism of \textit{second quantization}, as  the most convenient for our purposes (compare \cite{Fetter Walecka}, chapter 1). We introduce electron creation (annihilation) operators $c_{\mathbf{i}}^{\dag}$ ($c_{\mathbf{j}}$) for lattice site $\mathbf{i}$ ($\mathbf{j}$), respectively.
They obey standard anti-commutation relations
\begin{eqnarray}
c_{\mathbf{j}}c_{\mathbf{i}}^{\dag} + c_{\mathbf{i}}^{\dag}c_{\mathbf{j}}= 
\delta_{\mathbf{i}\mathbf{j}}, ~~~~ c_{\mathbf{j}}c_{\mathbf{i}} + c_{\mathbf{i}}c_{\mathbf{j}}= 0,
\label{fermion anticommutation relations}
\end{eqnarray}
and are related to  the basis state vectors $|\mathbf{i} \rangle$ ($\langle \mathbf{i}|$) in the following way
\begin{eqnarray}
|\mathbf{i} \rangle = c^{\dag}_{\mathbf{i}}|0\rangle, ~~~~ \langle \mathbf{i}|  = \langle0|c_{\mathbf{i}},
\label{Wannier states IInd quant}
\end{eqnarray}
where $|0\rangle$ is a \textit{vacuum state}, spanning zero-particle sector $\mathcal{F}_0 \subset \mathcal{F}$.


The dynamics of the electron can be then described as follows: initially localized in the vicinity of a particular site $\mathbf{i}$, it may quantum-mechanically  \textit{tunnel} to  neighboring site $\mathbf{j}$. In  $\mathcal{H}$ it corresponds simply to the transition from basis state $|\mathbf{i} \rangle$   to state $|\mathbf{j} \rangle$. If we include only the so-called \textit{single-band dynamics} and neglect other possible degrees of freedom (like spin, orbital, etc.) of the electron,  the resulting Hamilton operator (\textit{Hamiltonian}) reads
\begin{equation}
\hat{H} = \sum_{\mathbf{i}, \mathbf{j}  \in \mathbf{\Xi}}t(\mathbf{i}, \mathbf{j})c_{\mathbf{i}}^{\dag} c_{\mathbf{j}}  = \sum_{\mathbf{i}  \in \mathbf{\Xi}, \mathbf{q} \in  \mathcal{S}}t(\mathbf{q})c_{\mathbf{i}}^{\dag} c_{\mathbf{i + q}}.
\label{Lattice_Hamiltonian_General_ij_IIquatnization}
\end{equation}
Here  the summation over $\mathbf{i}$ runs through the whole lattice $\mathbf{\Xi}$. We assume 
that the complex numbers $t(\mathbf{i}, \mathbf{j})$ (\textit{hopping integrals}) fulfill the  following conditions,
\begin{equation}
\forall_{\mathbf{i},\mathbf{j} \in \mathbf{\Xi}}: ~~~t(\mathbf{i}, \mathbf{j}) = 
t(\mathbf{j}, \mathbf{i})^{\ast} = t(\mathbf{j}- \mathbf{i}) \equiv t( \mathbf{q}), ~~~~~  \mathbf{q} \equiv \mathbf{j}-\mathbf{i},
\label{conditions on ts}
\end{equation}
due to the fact that Hamiltonian is a hermitian operator, and because  all lattice sites of a given sub-lattice are assumed to be equivalent (\textit{translational invariance}). We also put $t(\mathbf{i}, \mathbf{i})=0$ for all $\mathbf{i}$.
By $\mathcal{S}$ we denote a set of vectors $\mathbf{q} $, for which $ t(\mathbf{q})\neq 0$. We assume the maximal number of independent hopping integrals $ t(\mathbf{q}) \equiv t_s \in \{t_1, t_2, \ldots, t_C\}$,
namely $C \equiv \frac{1}{2}|\mathcal{S}|$, the factor $\frac{1}{2}$ appears due to (\ref{conditions on ts}).   
Consequently, in general, the Hamiltonian does not possess all symmetry properties of the lattice - the discrete rotational symmetry may be lost unless we impose specific conditions on  parameters $t_s$. 

Due to translational invariance of the Hamiltonian (\ref{Lattice_Hamiltonian_General_ij_IIquatnization}), it is advisable to express it in \textit{Bloch basis}, labeled by the values of \textit{quasimomentum} $\mathbf{k}$, and related  
to the Wannier  basis by the unitary transformation,
\begin{equation}
c^{\dag}_{ \mathbf{k}}|0\rangle \equiv|\mathbf{k} \rangle = \frac{1}{\sqrt{\Lambda}}\sum_{\mathbf{i} \in \mathbf{\Xi}} \exp \big(-i\mathbf{k} \cdot  \mathbf{R} (\mathbf{i}) \big) |\mathbf{i} \rangle
\label{DFT i to k}
\end{equation}
with the inverse given by
\begin{equation}
c^{\dag}_{\mathbf{i}}|0\rangle \equiv|\mathbf{i} \rangle =  \frac{1}{\sqrt{\Lambda}}\sum_{\mathbf{k}\in FBZ} \exp(i\mathbf{k} \cdot \mathbf{R} (\mathbf{i}) ) |\mathbf{k} \rangle.
\label{DFT k to i}
\end{equation}
The above formulas  refer only to the case of finite  ($\Lambda < \infty$),  Bravais lattice ($M=1$). The formalism for non-Bravais lattices is not presented here, the limit  $\Lambda \to \infty$ will be considered later. 

The vectors $\mathbf{k}$ in (\ref{DFT i to k},\ref{DFT k to i}), determined by the lattice geometry and PBC imposed, belong to the \textit{first Brillouin zone} (FBZ), which is the \textit{Wigner-Seitz unit cell} of the reciprocal lattice centered at $\mathbf{K}=0$   (compare \cite{Kittel}, chapter 9).
 
Formulas (\ref{Wannier states IInd quant}), (\ref{DFT i to k}) and  (\ref{DFT k to i}) give us also the transformation properties of the  creation (annihilation) operators themselves, and can easily find the form of the Hamiltonian  (\ref{Lattice_Hamiltonian_General_ij_IIquatnization}) in the Bloch basis\footnote{Referred to also as the \textit{quasimomentum representation} or  $\mathbf{k}$-\textit{representation}.}, in which 
it  is diagonal, $\hat{H}|\mathbf{k} \rangle = \epsilon(\mathbf{k})|\mathbf{k} \rangle  $. Explicitly, we have
\begin{equation}
 \hat{H} = \sum_{\mathbf{k}} \epsilon(\mathbf{k})c^{\dag}_{\mathbf{k}}c_{\mathbf{k}} = \sum_{\mathbf{k}} \epsilon(\mathbf{k})\hat{n}_{\mathbf{k}}.
\label{general H in k}
\end{equation}
The functional dependence of its eigenvalues $\epsilon(\mathbf{k})$ on $\mathbf{k}$, the  \textit{ dispersion relation}, results from the lattice geometry, but is also a function  of  the hopping integrals $t_{s}$, $\epsilon = \epsilon(\mathbf{k}; t_1, \ldots, t_C)$. For $M=1$ it has  the following general form
\begin{equation}
\epsilon(\mathbf{k}; t_1, \ldots, t_C) =  t_1\epsilon_1(\mathbf{k}) + \ldots + t_C\epsilon_C(\mathbf{k}).
\label{epsilon }
\end{equation}
We assume, that $\epsilon(\mathbf{k}; t_1, \ldots, t_C)$ is at least a piecewise continuous, and then integrable,  function of $\mathbf{k}$.

Consider now a limit $\Lambda \to \infty$.
Due to the periodic boundary conditions (PBC), for any finite $\Lambda$ the number of $\mathbf{k}$ vectors in FBZ is equal that of lattice vectors $\mathbf{R}$, $\mathbf{R} \in \Lambda$. As a consequence, in the  limit in question, 
 some physical quantities, expressed as sums over $\mathbf{k}\in \text{FBZ}$, may be  divergent. To keep the values  of those quantities finite, we rescale them by the factor $\Lambda$, 
\begin{equation}
\sum_{\mathbf{k} \in FBZ} f(\mathbf{k} ) \longrightarrow \frac{1}{\Lambda} \sum_{\mathbf{k}\in FBZ}
 f(\mathbf{k}). 
\label{sum over k tends to infinity}
\end{equation}
The sum on the r.h.s of (\ref{sum over k tends to infinity}), for a given value of $\Lambda$, may be regarded as the Riemann sum of a function $f(\mathbf{k})$. The important fact is that a mesh of any partition of FBZ given by the values of  $\mathbf{k}$ goes to zero 
 in the limit $\Lambda \to \infty$, regardless of the particular choice of PBC.  Consequently, in that limit, (\ref{sum over k tends to infinity}) approaches the value of the  Riemann integral  over a FBZ of  $f(\mathbf{k})$, here understood as a function of continuous variable $\mathbf{k}$. We can then
replace  the summation in  (\ref{sum over k tends to infinity}) by the integration over FBZ, according to 
\begin{equation}
\frac{1}{\Lambda}\sum_{\mathbf{k} \in FBZ} f(\mathbf{k}) \buildrel\Lambda \to \infty \over\longrightarrow  \frac{1}{V } \int_{FBZ} f(\mathbf{k}) d\mathbf{k}, 
\label{sum to integral over k}
\end{equation}
where $ V $ is a Jordan  measure  of the first Brillouin zone. 


\subsection{The partition function}
\subsubsection{General formalism and quasimomentum representation.} 
According to quantum statistical mechanics (see e.g. \cite{statistical mechanics}, p. 245), the \textit{partition function}, $Z$, is given by the formula 
\begin{equation}
Z=Z(\beta) = \text{Tr} \big(\exp(- \beta \hat{H})\big),
\label{Z}
\end{equation}
where Tr denotes trace of an operator, $\beta =1/ T$ is an inverse temperature and $\hat{H}$ is the  Hamiltonian of a  system considered.
We evaluate $Z$  first in Bloch basis, which is rather simple, due to the fact that one-electron Bloch states $c^{\dag}_{\mathbf{k}}|0\rangle $ are eigenstates of the Hamiltonian (\ref{Lattice_Hamiltonian_General_ij_IIquatnization}). In that case the partition function reads 
\begin{equation}
Z=  \frac{1}{\Lambda} \sum_{FBZ} \exp(- \beta \epsilon(\mathbf{k})), ~~~ \Lambda < \infty,
\label{general Z in k-space discrete}
\end{equation}
for finite number of lattice vectors, and
\begin{equation}
Z=  \frac{1}{V}\int_{FBZ} \exp(- \beta \epsilon(\mathbf{k})) d\mathbf{k}
\label{general Z in k-space}
\end{equation}
for the  $\Lambda \to \infty$ limit.
To avoid infinite value of $Z$ in (\ref{general Z in k-space discrete}) we have applied normalization described in a previous paragraph by formula (\ref{sum over k tends to infinity}).
Introducing for convenience $C$ new variables $\xi_s$, defined as follows 
\begin{equation}
\xi_s = -\beta t_s, ~~~ s \in \{1, 2, \ldots, C\},
\label{beta t xi}
\end{equation}
and having in mind (\ref{epsilon }), we can rewrite both (\ref{general Z in k-space discrete}) and  (\ref{general Z in k-space}) as
\begin{equation}
Z = Z(\xi_1, \xi_2, \ldots, \xi_C)= \frac{1}{\Lambda} \sum_{FBZ} \exp \big(\sum_{1}^{C} \xi_s\epsilon_s(\mathbf{k})\big)
\label{general Z in k-space with ksi descrete}
\end{equation}
and
\begin{equation}
Z = Z(\xi_1, \xi_2, \ldots, \xi_C)= \frac{1}{V} \int_{FBZ} \exp \big(\sum_{1}^{C} \xi_s\epsilon_s(\mathbf{k})\big)d\mathbf{k}.
\label{general Z in k-space with ksi}
\end{equation}

Let us point out, that for most lattice geometries considered, the sum (\ref{general Z in k-space with ksi descrete}) or the integral 
(\ref{general Z in k-space with ksi})  cannot be easily evaluated, and the explicit form of the functional dependence  of $Z(\xi_1, \xi_2, \ldots, \xi_C)$ on its arguments cannot be evaluated either  in terms  of elementary functions or in terms of standard special functions. 
\subsubsection{Coordinate representation}

The general formula for the  partition function of our one-electron system, evaluated with the help of Bloch basis states (\ref{general Z in k-space with ksi descrete}), (\ref{general Z in k-space with ksi}), has a form of a complicated sum or integral, which usually cannot be  handled analytically. However, the trace of an operator does not depend on the basis one uses, and in order   to obtain an alternative expression for $Z(\xi_1, \xi_2, \ldots, \xi_C)$  we compute it now using Wannier basis.

Evaluation of the partition function in the present case requires some effort, below we describe a method used for this purpose.
First, assuming $\Lambda < \infty$, we rescale $Z$ by $\Lambda$, in correspondence with the formula (\ref{sum over k tends to infinity}), 
\begin{equation}
Z = \frac{1}{\Lambda}\sum_{\mathbf{i}\in \mathbf{\Xi}}  \langle \mathbf{i}| \exp(- \beta \hat{H}) | \mathbf{i}\rangle \ = \frac{1}{\Lambda}\sum_{\mathbf{i} \in \mathbf{\Xi}} \sum_{n}  \langle \mathbf{i}| \frac{(-\beta \hat{H})^n}{n!}   | \mathbf{i}\rangle.
\label{Z in Wannier rep}
\end{equation}
The problem in question is solved, once we know how to evaluate\footnote{We postpone for a moment a question if it is legitimate  to  interchange the  summation over the order of the expansion $n$ and that over the states $|\mathbf{i}\rangle$.} the quantity $g(n)$ defined as 
\begin{equation}
\frac{1}{\Lambda}\sum_{\mathbf{i}\in \mathbf{\Xi}}  \langle \mathbf{i}| (-\beta \hat{H})^n   | \mathbf{i}\rangle \ \equiv g(n),
\label{definition of g(n)}
\end{equation}
for the Hamiltonian (\ref{Lattice_Hamiltonian_General_ij_IIquatnization}). Explicitly, in this particular case we have
\begin{equation}
 g(n) = \frac{1}{\Lambda} \sum_{\mathbf{l}\in \mathbf{\Xi}}  \langle \mathbf{l}| \big(\sum_{\mathbf{i}, \mathbf{j(i)}\in \mathbf{\Xi}} -\beta t_{\mathbf{i} \mathbf{j}} A_{\mathbf{ij}} \big)^n |\mathbf{l}\rangle, 
\end{equation}
where we put $t(\mathbf{i} \mathbf{j}) \equiv t_{\mathbf{i} \mathbf{j}} $  and  introduce a shorthand $ A_{\mathbf{ij}} \equiv  c_{\mathbf{i}}^{\dag} c_{\mathbf{j}}$ to make the following formulas  more compact and transparent. 
One can easily examine some obvious properties of just introduced operators $ A_{\mathbf{ij}}$, below we invoke only two of them, 
\begin{equation}
A^{\dag}_{\mathbf{ij}} =  A_{\mathbf{ji}},~~~  
 A_{\mathbf{ij}}A_{\mathbf{ji}} = \hat{n}_{\mathbf{i}}(1 - \hat{n}_{\mathbf{j}}).
\label{two properties of Aij operators}
\end{equation}

When calculating $g(n)$ we have to deal with the terms of the  general form
\begin{equation}
(-\beta)^n t_{\mathbf{l_{2n} l_{2n-1}}}t_{\mathbf{l_{2n-2} l_{2n-3}}}\cdots  t_{\mathbf{l_4l_3 }} t_{\mathbf{l_2    l_1}}A_{\mathbf{l_{2n} l_{2n-1}}} A_{\mathbf{l_{2n-2} l_{2n-3}}}\cdots  A_{\mathbf{l_4l_3 }} A_{\mathbf{l_2    l_1}}, 
\label{general A term with hopping}
\end{equation}
where  $\mathbf{l_m} \in \mathbf{\Xi}$. In what follows we  concentrate on the operator part of (\ref{general A term with hopping}), 
\begin{equation}
A_{\mathbf{l_{2n} l_{2n-1}}} A_{\mathbf{l_{2n-2} l_{2n-3}}}\cdots  A_{\mathbf{l_4l_3 }} A_{\mathbf{l_2    l_1}}.
\end{equation}
Most of such terms do not contribute to $g(n)$. This is, firstly, due to the fermion anticommutation relations and properties of the  occupation number operator\footnote{More precisely, we refer to $\hat{n}_{\mathbf{i}}$ as to an \textit{electron} and  $1 -\hat{n}_{\mathbf{i}}$ as to a \textit{hole} number operator for a given site $\mathbf{i}$, respectively.}, that all sequences of the type 
\begin{equation}
\cdots A_{\mathbf{l i}} A_{\mathbf{l j}} \cdots, ~~~ \cdots A_{\mathbf{i l}} A_{\mathbf{ jl}} \cdots~~~ \text{or}~~~ \cdots  \hat{n}_{\mathbf{i}}(1 -  \hat{n}_{\mathbf{i}})  \cdots,
\label{types of Aij vanishing combinations}
\end{equation}
identically vanish. Secondly, because we compute trace in basis  of one-electron states, all terms where the product of two or more occupation electron number operators for different sites, $\mathbf{i}, \mathbf{j}$ appear, 
\begin{equation}
\cdots A_{\mathbf{i l}} A_{\mathbf{ li}} A_{\mathbf{j m}} A_{\mathbf{ m j}}\cdots = \cdots \hat{n}_{\mathbf{i}}(1 -  \hat{n}_{\mathbf{l}}) \hat{n}_{\mathbf{j}}(1 -  \hat{n}_{\mathbf{m}})\cdots,
\end{equation}
must vanish, too. This restricts the form of the terms in question to 
\begin{equation}
A_{\mathbf{i_{n+1} i_n}}A_{\mathbf{i_n i_{n-1}}}\cdots  A_{\mathbf{i_3 i_2}} A_{\mathbf{i_2 i_1}} = c_{\mathbf{i_{n+1}}}^{\dag} (1-\hat{n}_{\mathbf{i_n}})(1-\hat{n}_{\mathbf{i_{n-1}}})\cdots
(1-\hat{n}_{\mathbf{i_{2}}})c_{\mathbf{i_{1}}}.
\label{general nonzero A term nondiag}
\end{equation}
However, due to the trace operation we have to consider only diagonal terms, i.e. those with  the first and the last index in any product of $n$ $A_{\mathbf{ ij}}$ operators being identical. Consequently, the only allowed terms in an expansion giving nonzero contribution to $g(n)$ must be of the following form 
\begin{equation}
A_{\mathbf{i_{1} i_n}}A_{\mathbf{i_n i_{n-1}}}\cdots  A_{\mathbf{i_3 i_2}} A_{\mathbf{i_2 i_1}} = c_{\mathbf{i_{1}}}^{\dag} (1-\hat{n}_{\mathbf{i_n}})(1-\hat{n}_{\mathbf{i_{n-1}}})\cdots
(1-\hat{n}_{\mathbf{i_{2}}})c_{\mathbf{i_{1}}}.
\label{general nonzero A term}
\end{equation}
Notice that the above expression  is a product of $n-1$ hole number operators sandwiched in between $ c_{\mathbf{i_{1}}}^{\dag}$ and $ c_{\mathbf{i_{1}}}$ operators,
and has trace equal unity. 

Now let us  now return to the full form  of the terms (\ref{general nonzero A term}), 
\begin{equation}
(-\beta)^n t_{\mathbf{i_{1} i_n}} t_{\mathbf{i_n i_{n-1}}}\cdots t_{\mathbf{i_3 i_2}}t_{\mathbf{i_2 i_1}} A_{\mathbf{i_{1} i_n}}A_{\mathbf{i_n i_{n-1}}}\cdots  A_{\mathbf{i_3 i_2}} A_{\mathbf{i_2 i_1}},
\label{Final term A with t} 
\end{equation}
and assume for a moment that among $n$ hopping integrals in (\ref{Final term A with t}), $t_s$  appears  $n_s$ times, $s=1, 2, \ldots, C$. 
Having in mind  variables $\xi_{s}$ introduced in (\ref{beta t xi}), we see, that  (\ref{Final term A with t})  contributes to $g(n)$ given by (\ref{definition of g(n)}) through the term 
\begin{equation}
\xi_{1}^{n_1}\xi_{2}^{n_2}\cdots \xi_{C}^{n_C},~~~~~~~~~  \sum_{1}^C n_s = n. 
\label{xi ns term}
\end{equation}
We denote the total number of terms contributing in the same way to $g(n)$ by $\Gamma(n_1,  n_2, \ldots, n_C )$. 
To obtain $g(n)$ we must sum up the terms (\ref{xi ns term}) for all permitted values $n_1,  n_2, \ldots, n_C $,  
\begin{equation}
g(n; \xi_1, \xi_2,\ldots, \xi_C)  = \sum^{constr}_{n_1,  n_2, \ldots, n_C }  \Gamma(n_1,  n_2, \ldots, n_C )\cdot \xi_{1}^{n_1}\xi_{2}^{n_2}\cdots \xi_{C}^{n_C},
\label{sum of gamma contributions to g(n)} 
\end{equation}
where we have explicitly denoted the dependence of $g(n)$ on $ \xi_1, \xi_2, \ldots, \xi_C$, and the fact that  summation in (\ref{sum of gamma contributions to g(n)}) obeys the constraint appearing in (\ref{xi ns term}).
Comparing (\ref{Z in Wannier rep}), (\ref{definition of g(n)}) and (\ref{sum of gamma contributions to g(n)}), we have
\begin{equation}
Z(\xi_1, \xi_2,\ldots, \xi_C) = \sum_n  \sum^{constr}_{n_1,  n_2, \ldots, n_C } \frac{\Gamma(n_1,  n_2, \ldots, n_C )}{n!}\cdot \xi_{1}^{n_1}\xi_{2}^{n_2}\cdots \xi_{C}^{n_C}.
\label{Z through g(n)} 
\end{equation}
Obviously, from the above formula it follows that the $\Gamma(n_1,  n_2, \ldots, n_C )$ coefficients  are related to the respective coefficients $Z_{n_1  n_2 \ldots n_C}$ of the  Taylor series expansion of $Z(\xi_1, \xi_2,\ldots, \xi_C)$, 
\begin{equation}
\Gamma(n_1,  n_2, \ldots, n_C ) =  n! \cdot Z_{n_1  n_2 \ldots n_C},
\label{gamma vs expansion of Z} 
\end{equation}
defined as
\begin{equation}
Z_{n_1  n_2 \ldots n_C} =  \frac{\partial^{n_1}}{\partial \xi^{n_1}_1} \frac{\partial^{n_2}}{\partial \xi^{n_2}_2}  \cdots \frac{\partial^{n_C}}{\partial \xi^{n_C}_C} 
\frac{ Z(\xi_1,\ldots, \xi_C)}{n_1!n_2!\cdots n_C!}\Big|_{\xi_1=\xi_2 = \ldots =\xi_C =0}.
\label{Taylor coefficients of Z}
\end{equation}
However, according to the remark in section 2.1, in most cases (\ref{general Z in k-space with ksi descrete}), (\ref{general Z in k-space with ksi})) 
and their derivatives with respect to $\xi_s$ cannot be evaluated analytically. In order to find an explicit formula for the  Taylor series of the partition function, we have to develop an alternative method of computing  coefficients $Z_{n_1  n_2 \ldots n_C}$ (or $\Gamma(n_1,  n_2, \ldots, n_C )$). This problem is solved in the next chapter.
\subsubsection{The method of evaluating  $\Gamma(n_1,  n_2, \ldots, n_C )$.}
 There exist a method of combinatorial character, which allows us to find $ \Gamma(n_1,  n_2, \ldots, n_C )$ coefficients. It is based on the following observation:
the dynamics of our system, governed by the Hamiltonian (\ref{Lattice_Hamiltonian_General_ij_IIquatnization}) may be described by an intuitive picture of an electron tunneling - or  'taking a step' form one site to the other. More precisely, such a step is generated by $A_{\mathbf{ij}}$ operator, and may be visualized by an arrow,  pointing from  site $\mathbf{j}$ to site $\mathbf{i}$. Each such arrow is labeled by a complex number - a hopping integral $t_{\mathbf{ij}}= t_s$. 
Consequently, the product of operators given  by (\ref{general nonzero A term}) can be visualized, using terms of graph theory, as a \textit{closed path} or a  \textit{cycle},  with the respective lattice sites being its vertices, and with the complex number 
$t_s$ ascribed to each edge.
The crucial fact is the one-to-one correspondence between terms (\ref{general nonzero A term}) and the resulting closed paths, for details see appendix  \ref{appendix proof on otoc}.
For our purposes it is  important which vertex of our graph is a \textit{terminal} one, 
(i.e. this from which our path starts and where it ends), the  graphs differing by a terminal vertex are different from our point of view.
Notice also, that any vertex, including a terminal one, may be 'visited' by an electron more then once, i.e. there may be more than one edge connecting two vertices, and thus our closed paths does not have to be \textit{simple}. 

The set of all vertices that can be connected by a directed edge with a particular initial site are given by vectors $\mathbf{q} \in \mathcal{S}$. In other words, electron can move in direction and at the distances determined by those vectors, or equivalently,  by the hopping integrals 
$t(\mathbf{q})$. (This is the reason why it is convenient to assume the maximal number of distinct hopping integrals, i.e. one $t$ for  every pair $(\mathbf{q}, \mathbf{- q}), \pm \mathbf{q} \in \mathcal{S}$). 
The above described 
\textit{crystal directions}
should not be confused with $D$ mutually orthogonal \textit{spacial directions}, which, in principle can be chosen at will. Obviously, for the closed path we have to fulfill the condition that  a $D$-dimensional vector of the total displacement  must be equal zero. 

At first glance it seems than in order to find $\Gamma(n_1,  n_2, \ldots, n_C )$ for a given lattice geometry, we have to compute the total  number of the closed paths of length\footnote{By \textit{length} we mean, according to the terminology of graph theory, the number of edges, i.e. the order of the expansion  of $\exp(-\beta \hat{H})$ and not the distance in a real space.} $n$, with exactly $n_s$ edges labeled by the respective hopping integral $t_s$.
However, due to the translational invariance of the Hamiltonian and our normalization procedure, the  summation over terminal sites in trace operation  just cancel the normalization factor $1/\Lambda$.  What remains then is to compute the number of such paths with one particular site chosen as a terminal one,
\begin{eqnarray}
 &&\textit{Number of closed paths of length n, with $n_s$ edges} \nonumber \\ 
\Gamma(n_1,  n_2, \ldots, n_C) &=& \textit{ labeled by $t_s$, and with one particular site $\mathbf{i}$}\nonumber \\ 
&& \textit{chosen as the terminal vertex.}
\end{eqnarray}
This allows us to compute $Z(\xi_1, \xi_2, \ldots, \xi_C)$ also in the $\Lambda \to \infty$ limit, the difference with the case of finite lattice is that in the latter, when computing  $\Gamma(n_1,  n_2, \ldots, n_C )$, periodic boundary conditions must be taken into account.

The number of independent variables in $Z(\xi_1, \xi_2, \ldots, \xi_C)$ can be reduced from its maximal value $C$ by putting $\xi_s = \xi_w$ for some $ s, w$, after all coefficients $\Gamma(n_1,  n_2, \ldots, n_C )$ are found. 
In the extreme case, partition function becomes function of one variable $\xi$ only, 
and the total number of all  closed paths of the total length $n$, starting from a particular lattice site, denoted by $\tilde{\Gamma}(n)$, is equal
\begin{equation}
\tilde{\Gamma}(n) \equiv \sum^{\text{constr}}_{n_1,  n_2, \ldots, n_C }  \Gamma(n_1,  n_2, \ldots, n_C ) = g(n; 1, 1,\ldots, 1). 
\label{sum of gamma=1 contributions to g(n)} 
\end{equation}

To summarize the above discussion, we make the following observation. Define a graph $\gamma(\hat{H})$, such that its vertices are lattice sites of the lattice considered, and that  vertices $\mathbf{j}$ and $\mathbf{i}$ are connected by an edge if and only if $t_{\mathbf{ij}} \neq 0$. In the case of finite lattice and  for all  hopping integrals equal unity, $t_{\mathbf{ij}}= t=1$, the  matrix of the Hamiltonian in  the Wannier representation is  simply the \textit{adjacency matrix} of the graph $\gamma(\hat{H})$. This observation immediately shows the essence of our method, the $n$-th power of adjacency matrix contains information about the number of all paths of length $n$, its trace gives the total number of all closed paths. Generalization to the case of complex $t_s$ requires some kind of \textit{weighted graphs}. 

For infinite lattices such a simple picture is no longer available.
\subsubsection{The main formula} 
I the previous section we have presented the method of finding the coefficients $ \Gamma(n_1,  n_2, \ldots, n_C )$,  
which does not require explicit evaluation of the sum  (\ref{general Z in k-space with ksi descrete}) or  integral (\ref{general Z in k-space with ksi}) nor the respective partial derivatives. 
Equating two alternative expressions for $Z$, formulas  (\ref{general Z in k-space with ksi descrete}) or (\ref{general Z in k-space with ksi}) and  (\ref{Z through g(n)}), we obtain 
\begin{eqnarray}
 \frac{1}{\Lambda}\sum_{FBZ} \exp \big(\sum_{1}^{C} \xi_s\epsilon_s(\mathbf{k})\big) &=&  \sum_{n}  \sum^{\text{constr}}_{n_1,  n_2, \ldots, n_C }  \frac{\Gamma^{f}(n_1,  n_2, \ldots, n_C)}{n!} \xi_{1}^{n_1}\xi_{2}^{n_2}\cdots \xi_{C}^{n_C}, \nonumber \\
\label{key formula sum}
\end{eqnarray}
or, for the case of infinite lattice\footnote{Throughout the text we keep the same symbol $\Gamma$ for both $\Lambda < \infty$ and $\Lambda \to \infty$, only here  we distinguish both cases by respective superscript to avoid misunderstanding. }, 
\begin{eqnarray}
 \frac{1}{V}\int_{FBZ} \exp \big(\sum_{1}^{C} \xi_s\epsilon_s(\mathbf{k})\big)d\mathbf{k} &=&  \sum_{n}  \sum^{\text{constr}}_{n_1,  n_2, \ldots, n_C }  \frac{\Gamma^{i}(n_1,  n_2, \ldots, n_C )}{n!} \xi_{1}^{n_1}\xi_{2}^{n_2}\cdots \xi_{C}^{n_C}. \nonumber \\
\label{key formula}
\end{eqnarray}
This is the main result of the present paper. Above formulas, for particular cases of lattice geometries, generate mathematical identities that may be even yet unknown.

There are two points of view one can have on (\ref{key formula sum}), (\ref{key formula}). Firstly, its r.h.s is an explicit expression for the series expansion of $Z(\xi_1, \xi_2,\ldots, \xi_C)$ given by  the nontrivial sum or integral appearing on the l.h.s. Thus, it could be regarded as some kind of  generalization of the well-known formula (\ref{once less bessel}) giving a modified Bessel function $I_0(2\xi)$. Although the occurring series are rather complicated and may be found inconvenient for practical computational purposes, they may be interesting from the point of view of the mathematical analysis.

On the other hand, quasimomentum representation of  $Z(\xi_1, \xi_2,\ldots, \xi_C)$  may be used to obtain a coefficients $\Gamma(n_1,  n_2, \ldots, n_C )$ - numbers of closed paths of some specific kind, for a given lattice geometry. Namely, let us remind the Reader, that according to (\ref{Z through g(n)}), (\ref{gamma vs expansion of Z}) and (\ref{key formula}), (\ref{key formula sum}) we have 
\begin{equation}
\Gamma^{f}(n_1,  n_2, \ldots, n_C)  = \frac{n!}{n_1!n_2!\cdots n_C!}\cdot 
\frac{\partial^n \Big(   \frac{1}{\Lambda}\sum_{FBZ} \exp \big(\sum_{1}^{C} \xi_s\epsilon_s(\mathbf{k})\big)\Big)}{\partial \xi^{n_1}_1 \partial \xi^{n_2}_2 \cdots \partial \xi^{n_C}_C}, 
\label{number of graphs sum}
\end{equation}

\begin{equation}
\Gamma^{i}(n_1,  n_2, \ldots, n_C)  = \frac{n!}{n_1!n_2!\cdots n_C!}\cdot 
\frac{\partial^n \Big(\frac{1}{V} \int_{FBZ}  \exp \big(\sum_{1}^{C} \xi_s\epsilon_s(\mathbf{k})\big) d\mathbf{k} \Big)}{\partial \xi^{n_1}_1 \partial \xi^{n_2}_2 \cdots \partial \xi^{n_C}_C},
\label{number of graphs}
\end{equation}
for finite and infinite lattice, respectively. The r.h.s of the above formulas can be easily computed numerically, and 
therefore  our results may be interesting from the point of view of graph theory or combinatorics.

Finally, in the case of all $\xi_s$ equal, due to (\ref{sum of gamma=1 contributions to g(n)}) formula (\ref{key formula}) reads 
\begin{eqnarray}
Z(\xi) &= &\frac{1}{V}\int_{FBZ} \exp \big(\xi \sum_{1}^{C} \epsilon_s(\mathbf{k})\big)d\mathbf{k}= 
\sum_{n}   \frac{\tilde{\Gamma}(n)}{n!} \xi^{n}.
\label{key formula b}
\end{eqnarray}
In the next section the above equality is utilized frequently instead of the most general case (\ref{key formula}). 


\section{The results}

In this part we present our results. 
For each case we begin with the brief description of the geometry of the direct as well as the reciprocal lattice. Then,  the set $\mathcal{S}$, as well as $C$- the number of independent hopping integrals in the Hamiltonian (\ref{Lattice_Hamiltonian_General_ij_IIquatnization}),
the dispersion relation and the partition function in quasimomentum representation are given. 
Next, we explain in detail how to compute the partition function in Wannier representation, for each particular lattice geometry.
Finally, we compute explicitly first few terms of the Taylor series expansion of $Z$,  both by combinatorial techniques as well as by numerical integration of the partition function in $\mathbf{k}$-representation.

Except for the linear  chain lattice (see \ref{LI finite}), we consider infinite lattices only,
and most analyzed cases are Bravais lattices,  except for paragraph \ref{Bravais Lattices}  where we deal with the non-Bravais honeycomb and diamond lattices.
Real hopping integrals are assumed throughout the main text, the case of  complex integral for particular lattice geometry is analyzed in  appendix  \ref{complex t}. 



\subsection{Linear chain with nearest neighbor \\ hopping \label{LI}}
In the present paragraph we consider the linear chain of atoms, with the  Hamiltonian (\ref{Lattice_Hamiltonian_General_ij_IIquatnization}) given by  
\begin{equation}
\hat{H}_{\mathcal{L}} = t\sum_{i}\sum_{q = -1,1} c_{i}^{\dag} c_{i+q},
\label{Lattice_Hamiltonian_LC_t}
\end{equation}
i.e. with the only nonzero $t_s$ between nearest neighboring sites.
\subsubsection{The case of a finite lattice \label{LI finite}}
We begin from the case of  \textit{finite} number of sites,  consequently, we have to  impose the periodic boundary conditions (PBC).
Diagonalization of the Hamiltonian (\ref{Lattice_Hamiltonian_LC_t}) by Fourier transform yields a dispersion relation
\begin{equation}
\epsilon_k  =  2 t\cos \left( k \right),
\label{dispertion for n.n. l.c.}
\end{equation}
where  due to  the  PBC quasimomentum $k$ is given by 
\begin{equation}
k = \frac{2m\pi}{\Lambda},~~~~~ m\in \mathbb{Z}.
\label{linear chain k through m}
\end{equation}
To find $\Lambda$ allowed values of $m$ (or, equivalently, $k$), i.e. the  first Brillouin zone (FBZ), we consider separately the case of even and odd $\Lambda$. We obtain, respectively
\begin{equation}
m \in \{ -\frac{\Lambda}{2} + 1, \ldots,\frac{\Lambda}{2} - 1,\frac{\Lambda}{2} \},~~~~ \Lambda ~~ \text{even},  
\label{linear chain m Lambda even}
\end{equation}

\begin{equation}
m \in \{ -\frac{(\Lambda -1)}{2},  -\frac{(\Lambda -1)}{2} + 1, \ldots,\frac{(\Lambda -1)}{2} \},~~~~ \Lambda ~~  \text{odd}.  
\label{linear chain m Lambda odd}
\end{equation}
The resulting values of  $k$ are then 
\begin{equation}
k \in \{ -\frac{\pi(\Lambda - 2)}{\Lambda}, \ldots,   \frac{\pi (\Lambda - 2)}{\Lambda},     \pi \},~~~~ \Lambda ~~ \text{even}
\label{linear chain BZ even}
\end{equation}

\begin{equation}
k \in \{ -\frac{\pi(\Lambda - 1)}{\Lambda}, \ldots, \frac{\pi (\Lambda - 3)}{\Lambda}, \frac{\pi (\Lambda - 1)}{\Lambda}\},~~~~ \Lambda ~~ \text{odd}. 
\label{linear chain BZ odd}
\end{equation}
Obviously, the permitted values of $k$ are related to the $\Lambda$ roots of equation 
\begin{equation}
z^{\Lambda} = 1, ~~~~ z \in \mathbb{C}
\end{equation}
in  the following way 
\begin{equation}
k = \text{arg} z.
\end{equation}
The partition function  depends now on one variable, $Z_{ \mathcal{L}}\equiv Z_{ \mathcal{L}}(\xi)$,   $\xi = -\beta t$, expressed in quasimomentum representation (\ref{general Z in k-space}) it  reads 
\begin{equation}
Z_{ \mathcal{L}}(\xi, \Lambda) = \frac{1}{\Lambda}\sum_{k \in FBZ} e^{2\xi \cos(k)}.
\label{LC Z in k-space 1}
\end{equation}
with the FBZ given by (\ref{linear chain BZ even}, \ref{linear chain BZ odd}). 
We denoted explicitly the dependence of $Z_{ \mathcal{L}}$  on both its natural argument $\xi$ as well as on the number of lattice sites $\Lambda$.

In order to evaluate $Z_{ \mathcal{L}}(\xi, \Lambda)$ in the Wannier representation, notice, that from any site electron can move to the nearest neighboring site either in positive (clockwise) or negative (anti clockwise) direction along the chain. Denote number of such steps by $n^{+}$ and $ n^{-} $, respectively. Obviously,  $n = n^{+}+n^{-} $, we also define
\begin{equation}
d = d(n, \Lambda)=n^{+}-n^{-}.
\label{d for linear chain}
\end{equation}
For a closed path of lenght $n$ on the lattice with $ \Lambda$ sites, the following condition has to be fullfiled  
\begin{equation}
d(n, \Lambda) = c\Lambda,  ~~~~~~~ c \in \mathbb{Z}.
\label{d_od_c_i_Lambda}
\end{equation}
In order to find the permitted values of $c$, let us note, that for any $n$ and $\Lambda$ we have
\begin{equation}
n = w  \Lambda + r, ~~~~~ r < \Lambda.
\label{n}
\end{equation}
or, equivalently,
\begin{equation}
w \Lambda = n - n (\text{mod} \Lambda) = \Lambda \lfloor n/\Lambda \rfloor.
\label{w}
\end{equation}
Formula (\ref{n}) shows, that the electron can wind at most $w$ times around the chain, and then have still $r$ steps to make to complete the path. Consequently, $c$ can take any value between $-w$ and $w$,
\begin{equation}
c \in \{-w, -w + 1, \ldots, w - 1, w \} \equiv \mathcal{C}_{ \mathcal{L}},
\label{d_od_w_i_Lambda}
\end{equation}
but $r$ must be an even number, and for even $\Lambda $ only the even powers of $\xi$ are present in the expansion of $Z_{ \mathcal{L}}(\xi)$. 

The number of  closed paths  of lenght $n$ with exactly $n^{+}$ steps in the positive direction is ${ n \choose n^{+}} = n!/(n^{+}!n^{-}!)$, the  summation over all allowed values of $n^{+}$ yields  their total, 
\begin{equation}
\Gamma_{ \mathcal{L}}(n; \Lambda) \equiv \sum_{n^{+}} \frac{n!}{n^{+}!n^{-}!} = \sum_{d(n)}  \frac{n!}{ [ \frac{1}{2}(n+d(n))]! [ \frac{1}{2}(n-d(n))]! } 
\label{number_closed_paths_lin_chain_nn}
\end{equation}
due  to  (\ref{d for linear chain}) and where values of $d$ (or $n^{+}$) are given by  (\ref{d_od_c_i_Lambda}, \ref{d_od_w_i_Lambda}). 
Finally, the partition function reads
\begin{eqnarray}
 Z_{ \mathcal{L}}(\xi, \Lambda )& = &\sum_{n = 0}^{\infty} 
\frac{ \Gamma_{ \mathcal{L}}(n, \Lambda)}{n!}\xi^n  \nonumber= \sum_{n = 0}^{\infty} \frac{\xi^{n}}{n!} \sum_{d(n)} \frac{n!}{[ \frac{1}{2}(n+d)]![ \frac{1}{2}(n-d)]!} \\ &=& \sum_{n = 0}^{\infty}  \sum_{d(n)} \frac{\xi^{n}}{[ \frac{1}{2}(n+d)]![ \frac{1}{2}(n-d)]!} \equiv  \sum_{n = 0}^{\infty}  \mathcal{L}_{ n}(\Lambda) \xi^{n}.
\label{Z_L1_n_a}
\end{eqnarray}
Equating (\ref{Z_L1_n_a}) and  (\ref{LC Z in k-space 1}),  having in mind (\ref{linear chain k through m}), we obtain
\begin{equation}
Z_{ \mathcal{L}}(\xi, \Lambda) = \frac{1}{\Lambda}\sum_{m(\Lambda)} \exp\big(2\xi \cos(\frac{2m\pi}{\Lambda})\big)=  \sum_{n = 0}^{\infty} \mathcal{L}_{ n}(\Lambda)  \xi^{n}, 
\label{LC Z in k-space 2 equal in i-space}
\end{equation}
the values of $m(\Lambda)$ are given by (\ref{linear chain m Lambda even}) and   
(\ref{linear chain m Lambda odd}).
\subsubsection{The case of infinite lattice}
Consider now the limit  $\Lambda \to \infty$. Then no path can be  closed by moving around the chain. Consequently, only terms with $d = 0$ contribute to (\ref{number_closed_paths_lin_chain_nn}), (\ref{Z_L1_n_a}) and  $n$ must be even. Defining: 
$\nu= n/2$, we can write (\ref{Z_L1_n_a}) as
\begin{equation}
Z_{ \mathcal{L}}(\xi, \infty )\equiv Z_{ \mathcal{L}}(\xi) =  \sum_{n = 0, \text{even}}^{\infty}   \frac{\xi^{n}}{(n/2)!(n/2)!}  =  \sum_{\nu = 0}^{\infty}   \frac{\xi^{2 \nu}}{\nu!\nu!}.
\label{df}
\end{equation}
In the limit considered  $\text{FBZ} =[-\pi, \pi]$ and we replace the summation over FBZ by an integration (compare formula (\ref{sum to integral over k})). Equating two formulas for 
$Z_{ \mathcal{L}}(\xi)$, analogously to (\ref{LC Z in k-space 2 equal in i-space}) we find
\begin{equation}
\frac{1}{2\pi}\int_{-\pi}^{\pi}e^{2\xi \cos(k)}dk = \sum_{\nu = 0}^{\infty}   \frac{\xi^{2 \nu}}{\nu!\nu!}.
\label{Z integral linear chain  2 equal in i-space} 
\end{equation}
This is the equality (\ref{once less bessel}) quoted in the introduction, and $Z_{ \mathcal{L}}(\xi)=I_0(2\xi)$.


\subsection{Linear chain with nearest and next-nearest neighbor hopping}
Also in this paragraph we  analyze the case of an infinite linear chain, so the direct and reciprocal lattice geometry are the same as  in section 3.2.
However, now we assume slightly more general form of the Hamiltonian, with the nonzero values of the nearest- as well as  next-nearest-neighbor hopping integrals,
\begin{equation}
\hat{H}_{L} = \sum_{i}\sum_{q = -1,1} (t_1c_{i}^{\dag} c_{i+q} + t_2 c_{i}^{\dag} c_{i+2q}).
\label{Lattice Hamiltonian LC t1 and t2}
\end{equation}
This difference results in a change in  dispersion relation,
\begin{equation}
\epsilon_{L}(k) = 2t_1 \cos (k) + 2t_2 \cos (2k), ~~~~k  \in  [ -\pi , \pi ] \equiv \text{FBZ}. 
\label{epsilon_l2}
\end{equation}
In the present case the partition function depends on two variables $\xi_1=-\beta t_{1}$, $ \xi_2=-\beta t_{2}$. 
Written  in the quasimomentum representation  it reads
\begin{equation}
Z_{L}(\xi_1 ,\xi_2 )=\frac{1}{2 \pi}\int_{-\pi}^{\pi} e^{2\xi_1 \cos (k) + 2\xi_2 \cos (2k)} dk 
\equiv \sum_{n_1, n_2}L_{n_1, n_2} \xi_1^{n_1} \xi_2^{n_2}.
\label{Z_{L} integral and sum}
\end{equation}
Let us now turn to the problem of evaluating $Z_{L}(\xi_1 ,\xi_2 )$ in coordinate representation. 
We start with the simple observation that there are two types of 'steps', related to hopping integrals $t_1$ and $t_2$. Their  length in the real space, measured in lattice constants, is equal $1$ and $2$, respectively\footnote{The Reader is warned to not confuse the length of the 'step', $\lambda$,  with length of the path,  $n$, being the total number of steps.}. Let  $n_1$ and $n_2$ be the  numbers of corresponding steps, with $n = n_1 + n_2 $ being their total.
In contrary to the situation in paragraph 3.2, in the present case $n$ can be odd, but $n_1$ cannot, as it is impossible to close a path in such a case. Consequently, 
$n$ and $n_2$ must have the same parity, however for $n_1=0$,  $n_2 = n$ must be even. 

We denote the number of steps of the real-space length $\lambda$ in positive (negative) direction along the chain by $n_{\lambda}^{+}$ ($n_{\lambda}^{-}$) respectively. Obviously,  
\begin{equation}
n_{\lambda} = n_{\lambda}^{+} + n_{\lambda}^{-}, 
\label{n lambda LII}
\end{equation}
and we  also define
\begin{equation}
d_{\lambda} = n_{\lambda}^{+} - n_{\lambda}^{-}, ~~~ \lambda = 1, 2.
\label{d lambda LII}
\end{equation}
In the  case considered, closed paths are those, for which 
\begin{equation}
d_{1} + 2d_{2}= 0.
\label{d lambda LII constraint}
\end{equation}
To find  $\Gamma(n_1, n_2)$, 
let us first fix  $n^{+}_{1}, n^{+}_{2}$ on some values permitted by (\ref{d lambda LII constraint}).
The number of  closed paths with exactly $n^{+}_{1}, n^{+}_{2}$ 'positive' steps is 
\begin{equation}
{n \choose n_1 } {n_1 \choose n^{+}_1 }{n_2 \choose n^{+}_2 }  = \frac{n!}{n^{+}_1!n^{-}_1!n^{+}_2!n^{-}_2!}.
\label{Number of c.p. for L II 1}
\end{equation}
This is because we have to choose the order in which $n_1$  'short' ($\lambda = 1$), and $n-n_1$  'long' ($\lambda = 2$) steps appear in the path, and then choose those $n^{+}_{1}$ ($n^{+}_{2}$) of the $n_{1}$ ($n_{2}$) steps of each type, that are taken in the respective positive direction.
To obtain the $\Gamma_{L}(n_{1},n_{2})$ we have to sum (\ref{Number of c.p. for L II 1}) over allowed values of $n^{+}_{1}, n^{+}_{2}$. 

Before we do that, it is convenient to express  first $n^{+}_{\lambda}, n^{-}_{\lambda}$ through  $n_{\lambda}, d_{\lambda}$ with the help of the equations (\ref{n lambda LII}, \ref{d lambda LII}), and utilize the constraint (\ref{d lambda LII constraint}) to eliminate $d_1$. 
The only difficulty is that permitted values of $d_2$  depend in a nontrivial way on $n_1, n_2$. To find the explicit form of   this dependence, we have to consider two cases, $n_1 \geq 2n_2$ and $n_1 \leq 2n_2$. 
In the first case, for any value of $d_2$  ($|d_2|\leq n_2$), we are able to find such a corresponding  $d_1$ that the condition (\ref{d lambda LII constraint}) holds. 
In the case $n_1 \leq 2n_2$, value $d^{max}_2$ depends on $n_1$, and also  on the parity, $\mathcal{P}$, of both $\frac{1}{2} n_1$ and $n_2$, in the following way
\begin{equation}
 d^{max}_2 = \frac{1}{2}n_1 
~~\text{if}~~~~\mathcal{P}(\frac{n_1}{2})=\mathcal{P}(n_2), ~~~~~ \frac{1}{2}(n_1-2)~~~ \text{otherwise}. \nonumber
\end{equation}
Note, that for  $n_1=2n_2$ results for both cases  coincide.
Finally, we obtain the following formula for $\Gamma(n_1 n_2)$,
\begin{equation}
\Gamma(n_1, n_2) =
 \frac{(n_1 + n_2)!}{n_1!n_2!} \sum_{d_2} { n_1 \choose  \frac{1}{2}(n_1 - 2d_2)} { n_2 \choose  \frac{1}{2}(n_2 - d_2)},
\label{Total number of c.p. for L II 1}
\end{equation}
and after simple algebra, for the coefficients of the expansion (\ref{Z_{L} integral and sum}),
\begin{eqnarray}
L_{n_1, n_2} & =  & \sum_{d_2} \frac{1}{(\frac{1}{2}(n_1 + 2d_2))!(\frac{1}{2}(n_1 - 2d_2))!  (\frac{1}{2}(n_2 + d_2))!(\frac{1}{2}(n_2 - d_2))!}. ~~~~~~
\label{L n1 n2}
\end{eqnarray}
Summation in (\ref{L n1 n2}) is over the elements  of the set $ \mathcal{D}_2 $, 
\[
   \mathcal{D}_2 = \{ - d^{max}_2,   - d^{max}_2 + 2, \ldots, d^{max}_2- 2, d^{max}_2 \},
\]
and we recall once more the dependence of $d^{max}_2$ on $n_{1}, n_2$,
\begin{itemize}
\item for  $n_1 \leq 2n_2$ \\ $d^{max}_2 = \frac{1}{2}n_1$ 
~~if~ $\mathcal{P}(\frac{n_1}{2})=\mathcal{P}(n_2)$, ~~~ $\frac{1}{2}(n_1-2)$~ otherwise,
\item for $n_1 \geq 2n_2$ ~~~ $d^{max}_2 = n_2$.
\end{itemize}
Equation (\ref{Z_{L} integral and sum}), combined with the explicit form of $L_{n_1, n_2}$, (\ref{L n1 n2}), is a central result of this section, and may be regarded as generalization of (\ref{Z integral linear chain  2 equal in i-space}).

The coefficients $L_{n_1, n_2}$ have a quite complicated structure, however, there exist some  relations between them. This is because, due to identity 
\begin{equation}
\cos(2k)=2\cos^2(k)-1,
\end{equation}
$Z_{L}(\xi_1 ,\xi_2 )$ obeys the following partial differential equation
\begin{equation}
-\frac{\partial Z_{L}}{\partial \xi_2} + 
\frac{\partial^2 Z_{L}}{\partial \xi^2_1}- 2Z_{L} = 0. 
\end{equation}
This implies the relations between the coefficients (\ref{L n1 n2}),
\begin{equation}
(n_1+2)(n_1 + 1)L_{n_1+2, n_2} - (n_2 + 1)L_{n_1, n_2+1} -2L_{n_1, n_2}=0.
\end{equation}


\subsection{Triangular and bcc lattices}
In this section we examine two-dimensional triangular, as well as three- dimensional bcc lattices. The reason why we consider those two cases together is that  the formulas giving the number of closed paths have a very similar form for both lattices. This is due to the similar geometry of triangular and bcc lattices, with the  nearest neighbors of a given site forming two interpenetrating simplices in two (three) dimensions, respectively.   

\subsubsection{Triangular lattice}
Each site of an infinite triangular lattice has six nearest neighbors grouped in pairs along  three  lattice directions, their positions are given by $ \pm \mathbf{e}_i$, 
\begin{equation}
\mathbf{e}_1 =(1,0),~~~  \mathbf{e}_2=\frac{1}{2}(-1 ,  \sqrt{3}),~~~ \mathbf{e}_3 = -\frac{1}{2}(1 ,  \sqrt{3}).  
\end{equation} 
The reciprocal lattice of a triangular lattice is also a triangular one. The first Brillouin zone is a hexagon centered in the origin,  with the vertices  
\begin{equation}
\pm (2\pi, - \frac{2\pi\sqrt{3}}{3}),~~~\pm (2\pi, + \frac{2\pi\sqrt{3}}{3}),~~~  \pm (4\pi\sqrt{3}/3,0)),  
\label{hexagonal FBZ}
\end{equation} 
and area equal $8 \sqrt{3} \pi ^2  $. 
In the Hamiltonian (\ref{Lattice_Hamiltonian_General_ij_IIquatnization}) we assume  nonzero hopping integrals only  between the nearest neighbors  of a given site $\mathbf{i}$.  In general, we can relate a different hopping integral to each of the three lattice directions. 
Instead, we concentrate on the symmetric case $t_1 = t_2 = t_3 \equiv t$, there is however no conceptual difficulty to extend the following results to the most general one. The dispersion relation reads now
\begin{equation}
\epsilon_{T}(k_x, k_y) = 2t \cos (k_x)+4t \cos \left(\frac{k_x}{2}\right) \cos
   \left(\frac{\sqrt{3}k_y}{2}\right),
\end{equation}
and, consequently, 
\begin{equation}
Z_{T}(\xi) = \frac{1}{8 \sqrt{3} \pi ^2 } \int e^{\xi  \left(2 \cos (k_x)+4 \cos \left(\frac{k_x}{2}\right)
   \cos \left(\frac{\sqrt{3} k_y}{2}\right)\right)} d k_x dk_y = \sum_{n} T_n \xi^n. 
\label{Z triangular k}
\end{equation}
Now we pass to the problem of finding $Z_{T}(\xi)$ in coordinate representation. We denote the  number of steps in the direction given by $\pm \mathbf{e}_i$ as $\pm n_i$, with $ n^{+}_{i}+n^{-}_{i} \equiv n_{i}$,  and define 
\begin{equation}
d_i  = n^{+}_{i}-n^{-}_{i}, ~~~~ i = 1, 2, 3.
\label{d T}
\end{equation}
Let  $\Delta_x$ ($\Delta_y$) be the total displacement along $x$ ($y$) coordinate. We have then 
\begin{equation}
\Delta_x = d_1 - \frac{1}{2}(d_2 + d_3), ~~~~ \Delta_y = \frac{\sqrt{3}}{2}(d_2 - d_3),
\end{equation}
and, as for a closed paths we require  $\Delta_x$ = $\Delta_y= 0$, we obtain the condition  
\begin{equation}
d_1 = d_2 = d_3 \equiv d.
\label{d are equal}
\end{equation}
One can easily convince himself that maximal allowed value of $d$ is 
\begin{equation}
d_{max} = \min(n_1, n_2, n_3).
\label{d_{max}}
\end{equation}
The number of closed paths with exactly $n^{+}_i, n^{-}_i$ 'positive' and 'negative' steps in each direction is equal
\begin{equation}
\frac{n!}{n_1!n_2!n_3! } 
{ n_1 \choose n^{+}_1} { n_2 \choose n^{+}_2} { n_3 \choose n^{+}_3},
\label{Paths T}
\end{equation}
and the total number of all paths of length $n$ is then obtained  by summing first over all permitted values of $n^{+}_i$, and then over all values $n_i = n^{+}_i+ n^{-}_i$ obeying
\begin{equation}
n_1 + n_2 + n_3 = n.
\label{n const T}
\end{equation}
It is, however, convenient to eliminate first $n^{+}_i (n^{-}_i)$ with the help of (\ref{d T}), and apply the conditions (\ref{d are equal}), (\ref{d_{max}}).
As the result, the coefficients $T_n \equiv \tilde{\Gamma}(n)/n!$ are given by 
\begin{equation}
T_n  = \frac{1}{n!} \sum^{\text{constr}}_{n_1, n_2, n_3} \frac{n!}{n_1!n_2!n_3! } \sum_d { n_1 \choose \frac{1}{2}(n_1 + d)} { n_2 \choose \frac{1}{2}(n_2 + d)} { n_3 \choose \frac{1}{2}(n_3 + d)}.
\label{T_n first}
\end{equation}
The first summation in (\ref{T_n first}) is taken over those $n_1, n_2$ and $ n_3$ which satisfy  condition (\ref{n const T}). What is more, $n_1, n_2$ and $ n_3$ must have the same parity,  for even $n$ all $n_{i}$ are even, whereas for odd $n$ all $n_{i}$ 
must be odd, too. This reflects the fact, that any closed path can be regarded as combination of a number of elementary paths of the two kinds: first, with $n=2$, is a step forward and  step backward in some direction, the second, with $n=3$, consists of one step in each direction.  
The second summation runs over $d \in \{ - d_{max},   - d_{max} + 2, \ldots, d_{max}- 2, d_{max} \}$.
In  formula  (\ref{T_n first}) some factorials cancel, and  we finally obtain 
\begin{equation}
T_n =  \sum_{n_1, n_2, n_3}  \sum_d \prod_{i=1}^{3} \frac{1}{(\frac{1}{2}(n_i + d))!(\frac{1}{2}(n_i - d))!}.  
\end{equation}
Explicitly computing $Z_T(\xi)$ up to sixth order using the above formula, we obtain
\begin{equation}
Z_T(\xi)=1+3 \xi ^2+2 \xi^3+\frac{15 }{4}\xi^4+3 \xi^5+\frac{17}{6} \xi^6+ \mathcal{O}(\xi^7).
\end{equation}
This  coincides with the result obtained by integration of the expansion of the exponent in (\ref{Z triangular k}) to the same order in $\xi$.
\subsubsection{Bcc (body centered cubic) lattice}

This case is in  many respects  very similar to the just examined triangular lattice, with the main difference being higher dimensionality of the bcc lattice.
Each lattice site has now eight, instead of six, nearest neighbors, given by  vectors 
$\pm \mathbf{e}_i$,
\begin{eqnarray}
\mathbf{e}_1 =\frac{1}{2}(1,1,1),~~~\mathbf{e}_2 =\frac{1}{2}(-1,-1,1), \nonumber \\
\mathbf{e}_2 =\frac{1}{2}(-1,1,-1),~~~\mathbf{e}_4 =\frac{1}{2}(1,-1,-1). 
\end{eqnarray}   
Similarly to the case of triangular lattice, non-zero hopping integrals are assumed only between nearest neighbors, and we put all of them equal,  $t_1 = t_2 = t_3= t_4 \equiv t$ (consequently, $\xi_1=\xi_2 =\xi_3=\xi_4\equiv \xi$). Reciprocal lattice of the bcc lattice is the  fcc lattice (see \cite{Kittel}, p. 74), the first Brillouin zone is \textit{regular rhombic dodecahedron} with the volume $V = 16 \pi^3$.  
The  dispersion relation is 
\begin{equation}
\epsilon_{bcc}(k_x, k_y, k_z) = 8t \cos (\frac{k_x}{2})\cos (\frac{k_y}{2})\cos (\frac{k_z}{2}), 
\end{equation}
and the partition function in $\mathbf{k}$-representation reads
\begin{equation}
Z_{bcc}(\xi) = \frac{1}{16 \pi^3} \int e^{8 \xi  \cos (\frac{k_x}{2})\cos (\frac{k_y}{2})\cos (\frac{k_z}{2}) } d k_x dk_y dk_z \equiv \sum_{n} \mathcal{B}_n \xi^n. 
\label{Z bcc}
\end{equation}
The coefficients 
 $\mathcal{B}_n$ are obtained in a  way very similar to those for the triangular lattice, also  the notation is analogous and self-explanatory, we skip the derivation then and present only final results. $\mathcal{B}_n$ are given by
\begin{equation}
 \mathcal{B}_n = \frac{1}{n!} \sum_{n_1, n_2, n_3, n_4} \frac{n!}{n_1!n_2!n_3! n_4! } \sum_d \prod_{i=1}^{4}{n_i \choose \frac{1}{2}(n_i + d)} 
\label{B_n first}
\end{equation}
The first summation in (\ref{B_n first}) is taken  over those $n_1, n_2, n_3, n_4$ which  satisfy the condition $n_1 + n_2 + n_3+n_4 = n$, where $n_i$ is a number of steps in the direction given by $\pm \mathbf{e}_i$. Likewise in the previous section, all $n_{i}$ must have the same parity, which implies that  $n$ must be  even in the present case.
The second summation runs over the elements of the set 
\begin{equation}
\mathcal{D}_{bcc}= \{ - d_{max},   - d_{max} + 2, \ldots,  d_{max}- 2, d_{max} \}
\label{d bcc}
\end{equation}
where 
\begin{equation}
d_{max} = \min(n_1, n_2, n_3, n_4).
\end{equation}
After obvious  simplifications, we finally obtain
\begin{equation}
\mathcal{B}_n =  \sum_{n_1, n_2, n_3, n_4}  \sum_d \prod_{i=1}^{4} \frac{1}{(\frac{1}{2}(n_i + d))!(\frac{1}{2}(n_i - d))!}.  
\label{B_n final}
\end{equation}
Explicit evaluation of $Z_{bcc}(\xi)$ up to twelfth  order in $\xi$, either using the   formula (\ref{B_n final}) or by numerical integration of (\ref{Z bcc}), gives the  expansion
\begin{equation}
\mathcal{B}(\xi)= 1 + 4 \xi^2 + 9 \xi^4  + \frac{100}{9}\xi^6 + \frac{1225}{144}\xi^8 
+ \frac{441}{100}\xi^{10} + \frac{5929}{3600}\xi^{12} + \mathcal{O}(\xi^{14}).
\end{equation}
Interestingly,  for $n \leq 30$ all $\mathcal{B}_n$ are squares of rational numbers. The  question arises, is this true for all values of $n$? However, at the moment we are not able to prove this conjecture or to find counterexample.


\subsection{Honeycomb and diamond lattices \label{Bravais Lattices}}
Likewise in the previous section, here we also analyze together two lattices, namely the two-dimensional honeycomb (graphene) lattice and three-dimensional diamond lattice. Once again, the reason is the similar geometry of both lattices, in the present case three (four) nearest-neighbors of a given site form a two (three) dimensional simplex, respectively. 
This results in a very similar form of the Taylor series expansion for the partition function in this two cases. 
Both honeycomb and diamond are non-Bravais lattices, with the basis consisting of $M=2$ atoms, we denote 
the resulting two sublattices $A$ and $B$, respectively. As a consequence,  evaluation of the partition function in quasimomentum representation is not that straightforward as in the Bravais ($ M =1 $) case. However, this is a standard textbook problem, so we omit the detailed explanations. What is important, our method applies here without any serious modification.
\subsubsection{Honeycomb lattice}
The nearest neighbors of any lattice site $\mathbf{j} \in A$ of an infinite  honeycomb lattice are given by vectors
\begin{eqnarray}
\mathbf{e}_1 =\big(0,-\frac{\sqrt{3}}{3}\big),~~~~~\mathbf{e}_2 =\big(\frac{1}{2}, \frac{ \sqrt{3}}{6}\big),
~~~~~\mathbf{e}_3 =\big(-\frac{1}{2}, \frac{ \sqrt{3}}{6}\big).
\label{graphne neighbours}
\end{eqnarray}  
whereas the neighbors of site $\mathbf{i} \in B$ by $-\mathbf{e}_1, -\mathbf{e}_2, -\mathbf{e}_3$.
Because honeycomb lattice is, in fact, a triangular Bravais lattice with two-site atomic basis,  its reciprocal lattice  is also a triangular lattice, identical to that considered in section 5.1, due to the particular choice of vectors (\ref{graphne neighbours}). Consequently, also the first Brillouin zone is the same, i.e. it is a hexagon given by (\ref{hexagonal FBZ}).

I the Hamiltonian (\ref{Lattice_Hamiltonian_General_ij_IIquatnization}) the non-zero values of the hopping integrals are assumed only to three nearest neighbors of a given site, and we put them equal.
In the  present case the dispersion relation consist of $M=2$ subbands, 
\begin{equation}
\epsilon_{G, \sigma}(k_x, k_y) =  \sigma t \sqrt{3 + 2 \cos (k_x)+4 \cos \left(\frac{k_x}{2}\right) \cos  \left(\frac{\sqrt{3}k_y}{2}\right)},
\end{equation}
where $ \sigma = -1$ ($ \sigma = 1$)  refers to  lower (higher) subband, respectively. To evaluate the partition function we have to  integrate states within each subband, and than to add contributions from both of them. Then
\begin{eqnarray}
Z_{G}(\xi) &=& \frac{1 }{8 \sqrt{3} \pi ^2 } \int_{FBZ} \sum_{\sigma=-1,1} e^{\sigma\xi 
\sqrt{3 +2 \cos (k_x)+4 \cos \left(\frac{k_x}{2}\right) 
\cos \left(\frac{\sqrt{3} k_y}{2}\right)} } d k_x dk_y \nonumber \\ 
& = & \frac{1}{4 \sqrt{3} \pi ^2 }  \int_{FBZ} \cosh\big(\xi \epsilon_{G, 1}(k_x, k_y)\big)   d k_x dk_y \equiv \sum_{n} G_n \xi^n. 
\label{Z honeycomb k II}
\end{eqnarray}
Now let us evaluate $Z_{G}(\xi)$ in Wannier basis. We have $\Xi = M \Lambda = 2\Lambda$,   terminal vertex  $\mathbf{j}$ of any closed path may belong either to  sublattice $A$ or $B$. Those two cases are equivalent with respect to the way we compute the number of closed paths, the only consequence  is an additional factor $2$. 

In order to close the path, the number of steps in each of three lattice directions, taken form $A$ to $B$ must be equal to that of steps form $B$ to $A$, consequently a path must be of the form $A \to B \to A\to  \ldots \to B\to A$.
Denote the number of steps, taken from $\mathbf{j} \in A$  in the $i$-th direction (i.e. given by  $\pm \mathbf{e}_i$)  as $p_i$. 
To find $\tilde{\Gamma}(n)$
we have to choose those $p_i$ of the total $p$   steps $A \to B$ that are taken in $i$-th direction, and then  \textit{independently} do the same for steps $B \to A$. The resulting  number is equal  $\big(p!/(p_1!p_2!p_3!)\big)^2 $.
Summing over all values of $p_1, p_2, p_3$, which in the present case are of arbitrary parity, but  obey the  constraint $ p_1 + p_2 + p_3 = p  \equiv n/2$, we obtain  
\begin{equation}
\tilde{\Gamma}_G(n) = 2 \sum_{p_1, p_2, p_3} \Big(\frac{p!}{p_1!p_2!p_3! }\Big)^2 = n! G_n   
\label{G_n}
\end{equation}
where the factor $2$ was included according to the previous discussion.
Using formula (\ref{G_n}) we find the coefficients $G_n$ for $n\leq 6$. This gives us
\begin{equation}
G(\xi)=2+3 \xi^2+\frac{5}{4}\xi^4+\frac{31}{120}\xi^6 + \mathcal{O}(\xi^8).
\end{equation}
This is precisely what one gets expanding (\ref{Z honeycomb k II}) up to sixth order in $\xi$ and doing the remaining integrations.
\subsubsection{Diamond lattice}
Diamond lattice is related to the honeycomb lattice in very much the same way as 
bcc to the triangular one. It can be regarded as a fcc (face centered cubic) lattice with two-atom basis, and thus two sublattices, $A$ and $B$. Consequently, the first Brillouin zone is that of fcc lattice, namely \textit{truncated octahedron}, its volume is $V = 32 \pi^3$ (see \cite{Kittel}, p. 76, with $a=1$).
Each site of sublattice $A$ ($B$) has four neighbors, their positions are given by $\mathbf{e}_i$ ($-\mathbf{e}_i$),  respectively, where $\mathbf{e}_i$ vectors are  the following
\begin{eqnarray}
\mathbf{e}_1 =\frac{1}{4}(1,1,1),~~\mathbf{e}_2 =\frac{1}{4}(-1,-1,1),~~ \mathbf{e}_2 =\frac{1}{4}(-1,1,-1),~~\mathbf{e}_4 =\frac{1}{4}(1,-1,-1). \nonumber \\
\label{diamond qs}
\end{eqnarray}  
Likewise in the honeycomb lattice case, the dispersion relation consist of two subbands labeled by $\sigma = -1, 1$; it is related to the dispersion relation of the fcc lattice as follows,
\begin{equation}
\epsilon_{D\sigma}(k_x, k_y, k_z) = \sigma t \sqrt{4 + 4 \epsilon_{fcc}(k_x, k_y, k_z) }
\label{epsilon diamond k}
\end{equation}
where 
\begin{equation}
\epsilon_{fcc}(k_x, k_y, k_z) = 4t\Big(\cos \frac{k_x}{2} \cos
  \frac{k_y}{2}+ \cos \frac{k_x}{2} \cos
  \frac{k_z}{2}+\cos \frac{k_y}{2} \cos
  \frac{k_z}{2}\Big).
\end{equation}
The resulting partition function is then 
\begin{eqnarray}
Z_{D}(\xi) &=& \frac{1}{32 \pi^3} \int_{FBZ} \sum^1_{\sigma=-1} e^{\sigma\xi \sqrt{4 + 4 \epsilon_{fcc}(k_x, k_y, k_z) } } d k_x dk_y dk_z \equiv \sum_{n} D_n \xi^n   \nonumber \\ &=& \frac{1}{16 \pi^3} \int_{FBZ}  \cosh(\xi \sqrt{4 + 4 \epsilon_{fcc}(k_x, k_y, k_z) } \big) d k_x dk_y dk_z. 
\label{Z diamond k II}
\end{eqnarray}
The way we compute the number of closed paths in  the present case directly follows the case of honeycomb lattice, with the only difference that now we have three, instead of two, spatial dimensions. We have then
 \begin{equation}
\tilde{\Gamma}_{D}(n) = 2 \sum_{p_1, p_2, p_3, p_4} \Big(\frac{p!}{p_1!p_2!p_3!p_4! }\Big)^2 
= n! G_{n},
\label{G_n final}
\end{equation}
$n_i =  2 p_i,~ i=1, \ldots, 4$,  $ n=2p$, and  sum over $p_i$ is constrained,   $\sum_{i=1}^4 p_i = p$, but again each $p_i$ may be of arbitrary parity, and the factor $2$ appears due to the number of equivalent  sublattices.
$Z_{D}(\xi)$ expanded up to eighth order in $\xi$, reads explicitly
\begin{equation}
2 +4 \xi ^2 + \frac{7 }{3}\xi ^4+\frac{32 }{45}\xi ^6 + \frac{97 }{720}\xi^8 + \mathcal{O}(\xi^{10}).
\end{equation}



\section{Appendices}

\subsection{Appendix A. One to one correspondence between the terms given by the formula (\ref{general nonzero A term}) and respective closed paths \label{appendix proof on otoc}}

In this paragraph we give a simple justification of the fact, that there is a one to one correspondence between terms (\ref{general nonzero A term}) and closed paths they generate. 

Clearly, any  term (\ref{general nonzero A term}) leads to some  path. We have to show that   different such terms correspond to different paths. Let us consider some term in question, 
\begin{equation}
A_{\mathbf{i_{1} i_n}}A_{\mathbf{i_n i_{n-1}}} \cdots A_{\mathbf{i_{k+1} i_{k}}} A_{\mathbf{i_k i_{k-1}}} A_{\mathbf{i_{k-1} i_{k-2}}} \cdots 
A_{\mathbf{i_{j+1} i_{j}}} A_{\mathbf{i_j i_{j-1}}}A_{\mathbf{i_{j-1} i_{j-2}}} \cdots A_{\mathbf{i_2 i_1}}, 
\label{one to one proof 1}
\end{equation}
and  assume, that apart from the above, there exist some other ordering of its $n$ $A_{\mathbf{ij}}$ operators, which gives  the same closed path. 
Any such ordering may be obtained by some permutation of $A_{\mathbf{ij}}$ operators in formula (\ref{one to one proof 1}), it is enough then to examine the case of a transposition.
Let us swap $A_{\mathbf{i_k i_{k-1}}}$  and  $A_{\mathbf{i_j i_{j-1}}}$, assuming first that   $A_{\mathbf{i_j i_{j-1}}} \neq  A_{\mathbf{i_2 i_{1}}}$ and $A_{\mathbf{i_k i_{k-1}}} \neq  A_{\mathbf{i_1 i_{n}}}$,
\begin{equation}
A_{\mathbf{i_{1} i_n}}A_{\mathbf{i_n i_{n-1}}} \cdots A_{\mathbf{i_{k+1} i_{k}}} A_{\mathbf{i_j i_{j-1}}} A_{\mathbf{i_{k-1} i_{k-2}}} \cdots 
A_{\mathbf{i_{j+1} i_{j}}} A_{\mathbf{i_k i_{k-1}}}A_{\mathbf{i_{j-1} i_{j-2}}} \cdots A_{\mathbf{i_2 i_1}}. 
 \label{one to one proof 2}
\end{equation}
Obviously if   $A_{\mathbf{i_k i_{k-1}}} = A_{\mathbf{i_j i_{j-1}}}$, this transposition does not affect (\ref{one to one proof 1}), yielding identical term. In the opposite case, after writing some $A_{\mathbf{ij}}$ operators in terms of creation and annihilation operators, (\ref{one to one proof 2}) reads
\begin{equation}
A_{\mathbf{i_{1} i_n}}A_{\mathbf{i_n i_{n-1}}} \cdots A_{\mathbf{i_{k+1} i_{k}}}
 c^{\dag}_{\mathbf{i_j}} c_{\mathbf{ i_{j-1}}} A_{\mathbf{i_{k-1} i_{k-2}}} \cdots 
c^{\dag}_{\mathbf{i_{j+1}}}  c_{\mathbf{ i_{j}}} A_{\mathbf{i_k i_{k-1}}} 
A_{\mathbf{i_{j-1} i_{j-2}}}
 \cdots A_{\mathbf{i_2 i_1}}. 
 \label{one to one proof 3}
\end{equation}
There are two possibilities now: either some operators  in between $c^{\dag}_{\mathbf{i_j}} $ 
and $c_{\mathbf{i_j}} $ carry site index $\mathbf{i_j}$ or this index does not appear there. In the former case, after commuting some operators, a
part of the term  (\ref{one to one proof 2}) has the form	
$\cdots c^{\dag}_{\mathbf{i_j}} (1 - n_{\mathbf{i_j}})  c_{\mathbf{i_j}} \cdots = \cdots c^{\dag}_{\mathbf{i_j}}  c_{\mathbf{i_j}} \cdots= \cdots n_{\mathbf{i_j}} \cdots$,
in the latter case that part also reduces to $\cdots c^{\dag}_{\mathbf{i_j}}  c_{\mathbf{i_j}} \cdots  = \cdots n_{\mathbf{i_j}} \cdots $. Consequently, in both cases the whole term (\ref{one to one proof 2}) vanishes, due to the fact that we compute trace  using one-particle states only.
The same reasoning applies to the reversal of order in which $c^{\dag}_{\mathbf{i_k}}$ and $c_{\mathbf{i_k}}$ appear in (\ref{one to one proof 2}).

If one, but not both, of  the permuted operators is $ A_{\mathbf{i_2 i_{1}}}$ or  $ A_{\mathbf{i_1 i_{n}}}$ the above reasoning applies without a modification.
Finally, if we interchange $ A_{\mathbf{i_2 i_{1}}}$ with  
 $ A_{\mathbf{i_1 i_{n}}}$, in order to  obtain the non-zero term, 
$\mathbf{i_2}= \mathbf{i_n}$ must hold. But then we obtain the path with  terminal vertex  $\mathbf{i_2} \neq \mathbf{i_1}$, so different from our point of view. 


\subsection{Appendix B. Linear chain of atoms  with complex nearest-neighbor hopping integrals \label{complex t}}

Consider the case analyzed in section \ref{LI}, but  let the hopping integral be an arbitrary complex number, $t=|t|e^{i \varphi}$.
This leads to a modified dispersion relation,
\begin{equation}
\epsilon_{ \mathcal{L}}(k, \varphi)  =  2 |t|\cos \left( k + \varphi  \right).
\label{dispertion for n.n. l.c. complex}
\end{equation}
Consequently, the partition function is given by 
\begin{equation}
Z_{ \mathcal{L}}(\rho, \varphi, \Lambda)= \frac{1}{\Lambda}\sum_{k \in FBZ} 
\exp\big(- 2 \rho \cos(k + \varphi)\big),  
\label{LC Z in k-space 1 complex}
\end{equation}
where $ \xi = -\beta t  = -\rho e^{i \varphi}$ and $\rho = |\xi| $. $Z_{ \mathcal{L}}(\rho, \varphi, \Lambda)$ is obviously real, periodic function of $\varphi$ with a period $2\pi$. Expanded in a  Fourier series it reads
\begin{equation}
Z_{ \mathcal{L}}(\rho, \varphi, \Lambda) = \sum_{p = -\infty}^{\infty} C_p(\rho) e^{i p \varphi },~~~~~~ C^{\ast}_p(\rho)= C_{-p}(\rho).
\label{fourier l.c. varphi 1}
\end{equation}
On the other hand, from the way we compute  $Z_{ \mathcal{L}}$ in coordination representation, it is clear that each step in a positive (negative) direction  brings a factor $e^{i\varphi}$ ($e^{-i\varphi}$), respectively, and that each path is characterized by a  total phase $\varphi d$,
\begin{equation}
 Z_{ \mathcal{L}}(\rho, \varphi, \Lambda ) =  \sum_{n = 0}^{\infty} \sum_{d(n)} \frac{(-\rho)^{n} (e^{i  \varphi })^d}{[ \frac{1}{2}(n+d)]![ \frac{1}{2}(n-d)]!}.  
\label{Z_L1_n_a complex}
\end{equation}
This shows, that in the expansion (\ref{fourier l.c. varphi 1}) the only nonzero terms are those with $p = d = c\Lambda$, and $d$ given by (\ref{d_od_c_i_Lambda}, \ref{d_od_w_i_Lambda}).
Comparing (\ref{Z_L1_n_a complex}) and (\ref{fourier l.c. varphi 1}), we see that $C_d(\rho)= C_{-d}(\rho)$, and  that (\ref{fourier l.c. varphi 1}) can be rewritten as  
\begin{equation}
Z_{ \mathcal{L}}(\rho, \varphi, \Lambda) =  a_{0}(\rho)+ \sum_{d(n) > 0} a_d(\rho) 
\cos ( \varphi d),~~~~~~ 
\label{fourier l.c. varphi}
\end{equation}
where 
\begin{equation}
a_{0}(\rho) = \frac{1}{2 \pi \Lambda } \int_{-\pi}^{\pi} \sum_{k \in FBZ} 
e^{- 2 \rho \cos(k + \varphi)} d \varphi = \sum_{n=0}^{\infty} \frac{(-\rho)^n}{\big( (\frac{1}{2}n)!\big)^2},
\label{fourier l.c. varphi coefficient a_0}
\end{equation}

\begin{equation}
a_{d}(\rho) = \frac{1}{\pi  \Lambda} \int_{-\pi}^{\pi} \sum_{k \in FBZ} 
e^{- 2 \rho \cos(k + \varphi)}\cos(\varphi d) d \varphi = \sum_{n=0}^{\infty} \frac{2 (-\rho)^{n} }{[ \frac{1}{2}(n+d)]![ \frac{1}{2}(n-d)]!}. 
\label{fourier l.c. varphi coefficient a_d}
\end{equation}
Assume now even $\Lambda$,  then $n = 2 \nu $, $d = 2 \delta $. Equating (\ref{Z_L1_n_a complex}) to  (\ref{fourier l.c. varphi}) and having in mind (\ref{fourier l.c. varphi coefficient a_0}, \ref{fourier l.c. varphi coefficient a_d}) we obtain the following generalization of (\ref{LC Z in k-space 2 equal in i-space}),
\begin{eqnarray}
\sum_{k \in FBZ} \frac{ \exp\big(-2\rho \cos(k + \varphi)\big)}{\Lambda} &= & \sum_{\nu = 0}^{\infty}  \Big(   \frac{1}{\nu!\nu!} + 2 \sum_{\delta(\nu) > 0} \frac{  \cos( 2 \delta \varphi ) }{[ (\nu + \delta)]![\nu - \delta]!}  \Big)\rho^{2\nu} \nonumber \\
&= & \sum_{\nu = 0}^{\infty}  \Big( \sum_{\delta(\nu) } \frac{  \cos( 2 \delta \varphi ) }{[ (\nu + \delta)]![\nu - \delta]!}  \Big)\rho^{2\nu}. 
\label{LC Z in k-space 2 equal in i-space  complex even Lambda}
\end{eqnarray}
If we put  $\varphi = \pi$ in the above formula 
we  obtain the results given by  (\ref{Z_L1_n_a},\ref{LC Z in k-space 2 equal in i-space}). On the other hand, for 
$\varphi = \frac{\pi}{2}$ we have $\cos(2 \delta \varphi)= \cos(\delta \pi) = (-1)^{\delta}$ and then
\begin{equation}
\frac{1}{\Lambda}\sum_{k \in FBZ} \exp\big(2\rho  \sin(k)\big) =  
\sum_{\nu = 0}^{\infty}  \Big( \sum_{\delta(\nu)} \frac{  (-1)^{\delta} }{[ (\nu + \delta)]![\nu - \delta]!}  \Big)\rho^{2\nu}.
\label{LC Z in k-space 2 equal in i-space  complex phi half}
\end{equation}






\subsection*{Summary and conclusions}
In this paper we have presented the method of evaluating the partition function of a single-electron, periodic system, using specific basis, in which the system Hamiltonian is not diagonal. Utilizing the symmetry properties of the system and its Hamiltonian, as well as   invariance of trace operation with respect to change of the basis in the Hilbert space, we are able to establish several mathematical identities. They could be found important  from the point of view of both pure and computational mathematics.  

The possible extensions of the present works include first analysis of infinite lattices  not examined yet, like fcc or kagome lattice, as well as detailed investigation of finite lattices for geometries more complicated than that of linear chain, and also for various PBC. 
The extension of the method to many-electron systems or systems without translational invariance is also possible in principle, but the technical difficulties that arise make the application of  combinatorial techniques  very problematic in these cases.

\subsection*{Acknowledgments}
I'm grateful to Prof. J\'{o}zef Spa\l ek for  
 his comments, guidance and encouragement. The hospitality of  the Zaremba Association of Mathematicians - Students of the Jagiellonian University  is also acknowledged.

\end{document}